\documentclass[aps,prd,reprint,preprintnumbers,superscriptaddress,amsmath,amssymb,floatfix,longbibliography]{revtex4-2}
\pdfoutput=1
\usepackage{subfigure}
\usepackage{graphicx}
\usepackage{dcolumn}
\usepackage{amsmath}
\usepackage{bm}
\usepackage[colorlinks,citecolor=blue,urlcolor=blue,linkcolor=red]{hyperref}

\begin{document}
\title{Scalar and tensor structures in $J/\psi J/\psi$ scattering from lattice QCD}

\author{Geng Li}
\email{ligeng@ihep.ac.cn}
\affiliation{Institute of High Energy Physics, Chinese Academy of Sciences, Beijing 100049, People's Republic of China}
\affiliation{Center for High Energy Physics, Henan Academy of Sciences, Zhengzhou 450046, People's Republic of China}

\author{Chunjiang Shi}
\affiliation{Institute of High Energy Physics, Chinese Academy of Sciences, Beijing 100049, People's Republic of China}
\affiliation{School of Physical Sciences, University of Chinese Academy of Sciences, Beijing 100049, People's Republic of China}

\author{Ying Chen}
\email{cheny@ihep.ac.cn}
\affiliation{Institute of High Energy Physics, Chinese Academy of Sciences, Beijing 100049, People's Republic of China}
\affiliation{Center for High Energy Physics, Henan Academy of Sciences, Zhengzhou 450046, People's Republic of China}
\affiliation{School of Physical Sciences, University of Chinese Academy of Sciences, Beijing 100049, People's Republic of China}

\author{Wei Sun}
\affiliation{Institute of High Energy Physics, Chinese Academy of Sciences, Beijing 100049, People's Republic of China}

\begin{abstract}
The $^1S_0$ and $^5S_2$ amplitudes of $J/\psi~J/\psi$ scattering are determined up to 6.6~GeV from $N_f=2$ lattice QCD simulations at $m_\pi \approx 420$~MeV and $250$~MeV. 
We observe an attractive interaction in the ${}^1S_0$ channel, which permits the existence of a near-threshold scalar structure. 
This structure may correspond to $X(6200)$, but the possible left-hand-cut effect prevents us from a precise pole determination. 
In the ${}^5S_2$ channel, while a near-threshold repulsive interaction is observed, a resonance appears at higher energies associated with a Castillejo-Dalitz-Dyson zero $\sqrt{s}_{\rm CDD}\approx 6.45~\rm{GeV}$ of the scattering amplitude. 
The tensor resonance has the parameters  $(m_R,\Gamma_R)=(6.543(10), 0.548(34))~\rm{GeV}$ for $m_\pi\approx 420~\rm{MeV}$ and $(6.538(13),0.537(56))~\rm{GeV}$ for $m_\pi\approx 250~\rm{MeV}$, which are compatible with those of  $X(6600)$ (or $X(6400)$) reported by ATLAS and CMS, and also support the $2^{++}$ assignment of the recent CMS study. 
The difference of the  $^1S_0$ and $^5S_2$ $J/\psi J/\psi$ interaction is attributed to the dominance of the quark rearrangement effects. 
The systematic uncertainties owing to the unphysical pion mass, the finite lattice spacing, and the finite volume should be investigated by future lattice QCD studies with more sophisticated lattice setups.
\end{abstract}

\maketitle

\textit{Introduction.}
$X(6900)$ [also labeled as $T_{cc\bar{c}\bar{c}}(6900)^0$ in PDG~2024~\cite{ParticleDataGroup:2024cfk}] was first observed by LHCb from the di-$J/\psi$ spectrum in 2020~\cite{LHCb:2020bwg} and was independently confirmed by
the ATLAS~\cite{ATLAS:2023bft} and CMS experiments~\cite{CMS:2023owd,Wang:2024koq}. 
Although $X(6900)$ is a promising candidate for the fully charmed tetraquark $T_{cc\bar{c}\bar{c}}$, most theoretical models predict a lighter $T_{cc\bar{c}\bar{c}}$ state below $X(6900)$.
Experimentally, the di-$J/\psi$ spectrum measured by LHCb, ATLAS, and CMS exhibits a broad structure in the 6.2–6.7~GeV range.
Assuming one or two resonances in this region, ATLAS analyzed the spectrum and reported a broad structure(s) referred to as $X(6600)$ [or $X(6400)$ and $X(6600)$]~\cite{ATLAS:2023bft}.
CMS also observed the broad structure $X(6600)$, along with an additional resonance $X(7100)$~\cite{CMS:2023owd,Wang:2024koq}.
Recently, CMS sorted $X(6600)$, $X(6900)$, and $X(7100)$ into a family of $T_{cc\bar{c}\bar{c}}$ states of the preferred $2^{++}$ quantum numbers~\cite{CMS:2025fpt, CMS:2025xwt}. On the other hand, a near-threshold structure $X(6200)$ was reported by phenomenological studies after analyzing the $J/\psi J/\psi$ spectra of LHCb, ATLAS, and CMS~\cite{Dong:2020nwy,Song:2024ykq}.

Theoretically, fully heavy tetraquark states were first investigated in the 1980s~\cite{Iwasaki:1975pv,Chao:1980dv,Ader:1981db,Li:1983ru,Heller:1985cb,Badalian:1985es}.
The QCD sum rule calculations~\cite{Chen:2016jxd,Wang:2017jtz,Wang:2018poa} predicted the fully charmed tetraquark spectra with various quantum numbers. Following the discovery of $X(6900)$ and its family, extensive theoretical studies were spurred~\cite{Guo:2020pvt,liu:2020eha,Chen:2020xwe,Maiani:2020pur,Dong:2020nwy,Chao:2020dml,Gong:2020bmg,Zhu:2020xni,Liang:2021fzr,Liu:2021rtn,Dong:2021lkh,Niu:2022jqp,Dong:2022sef,Wang:2022xja,Kuang:2023vac,Huang:2024jin,Wu:2024euj,Zhang:2024qkg} (see Ref.~\cite{Chen:2022asf} for a review). 
Since these $T_{cc\bar{c}\bar{c}}$ structures were observed in the di-$J/\psi$ system, it is of great interest to understand the dynamics between the two charmonium states that drive their formation.
Some phenomenological studies attribute the dynamics to pomeron or light meson exchanges between the two charmonia~\cite{Gong:2020bmg,Dong:2020nwy,Huang:2024jin}. 
Nevertheless, these theoretical interpretations generally depend on various underlying assumptions.

This Letter aims to explore the $J/\psi J/\psi$ interactions in the  ${}^1S_0$ ($0^{++}$) and ${}^5 S_2$ ($2^{++}$) channels via lattice QCD. In order to avoid the complication due to coupled-channel effects, we concentrate on the energy region below 6.6~GeV, where only the $\eta_c \eta_c$ and $J/\psi J/\psi$ channels are relevant. 
As addressed in the longer companion paper~\cite{Li:2025vbd}, the $J/\psi J/\psi$ channel is almost decoupled from the $\eta_c\eta_c$ channel, and thereby can be analyzed through the single-channel L\"{u}scher formalism~\cite{Luscher:1986pf,Luscher:1990ux,Luscher:1991cf}. 
We begin with the determination of the finite-volume energies (FVEs) of the $J/\psi J/\psi$ systems from the related correlation functions, through which the scattering amplitudes are extracted. 
Subsequently, we analyze the pole structure of the amplitudes and the dicharmonium interactions using appropriate parametrizations.

\textit{Finite volume energies.}
Our calculation is performed on anisotropic lattices with two degenerate light quark flavors.
The spatial lattice spacing is set to be $a_s\approx 0.136$~fm and the aspect ratio is tuned to be $\xi=a_s/a_t\approx 5.0$, where $a_t$ is the temporal lattice spacing~\cite{Li:2024pfg}. 
We perform calculations at two pion masses, $m_\pi\approx 420(250)$~MeV, labeled as M420 (M250).
The same fermion action for the valence charm quark is adopted, which is tuned to reproduce the spin-averaged $1S$ mass of $\left(M_{\eta_c} + 3M_{J/\psi} \right) / 4 = 3069$~MeV~\cite{Li:2024pfg}. 
For each $m_\pi$, we consider two volumes with spatial sizes $L_s=12a_s $ (denoted by L12) and $16a_s$ (denoted by L16). 
The lattice $J/\psi$ masses used in this work are $3.08850(86)$~GeV (L12) and $3.08714(48)$~GeV (L16) for M420, and $3.08166(91)$~GeV (L12) and $3.08528(45)$~GeV (L16) for M250.
The correlation functions of meson-meson operators are computed using the distillation method~\cite{Peardon:2009gh}. 

\begin{figure}[t]
	\centering
	\subfigure[Direct (D)]{
		\includegraphics[width=0.45\linewidth]{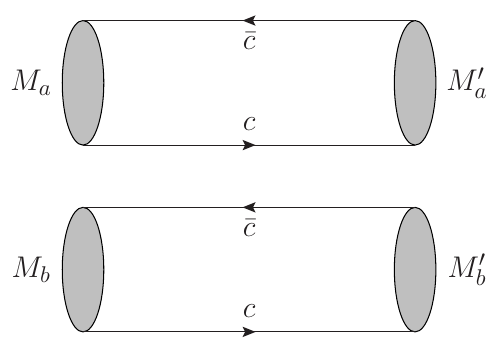}}
	\subfigure[Quark rearrangement (Q)]{
		\includegraphics[width=0.45\linewidth]{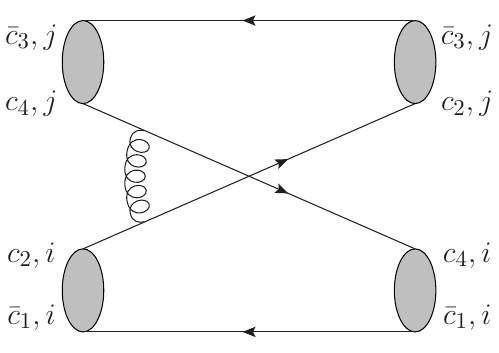}}
	\caption{Schematic illustrations of quark Wick contractions.}
	\label{fig:wick}
\end{figure}

We calculate FVEs of the ${}^1S_0$ and ${}^5S_2$ $J/\psi J/\psi$ systems up to a center-of-mass energy around 6.6~GeV in their rest frame. 
The details of the meson-meson operator construction can be found in the longer companion paper~\cite{Li:2025vbd}. 
Our $J/\psi J/\psi$ operators have the structure $\mathcal{O}_{\psi\psi}(n^2) = \mathcal{O}^i_\psi(\vec{k}) \circ \mathcal{O}^i_\psi(-\vec{k})$, where $\mathcal{O}_\psi(\vec{k})$ is the single particle operator projected to the lattice momentum $\vec{k}=(2\pi/L_s)\vec{n}$ with $n^2=|\vec{n}|^2$, and the symbol ``$\circ$" stands for the specific combinations of the different orientations of $\vec{k}$ and the indices of $\mathcal{O}_\psi^i$~\cite{Cheung:2017tnt}. 
The total spins $s=0$ and 2 are realized by combining the indices of $\mathcal{O}_\psi(\vec{k})$ to give the $A_1$ and $E$ representations of the cubic group O, respectively. 
The $S$-wave ($l=0$) is given by the $A_1$ representation of the spatial configuration of $\vec{k}$. 
Although the spatial $A_1$ contains both $l=0$ and $l\ge 4$ partial waves, the latter are tentatively neglected in this exploratory study. 
Note that the $S$-wave $\eta_c\eta_c$ operators should be included for the ${}^1S_0$ $J/\psi J/\psi$ FVEs to be determined unambiguously~\cite{Li:2025vbd}. 
In the practical calculation of the correlation matrices of these operators, we consider the two types of quark diagrams illustrated in Fig.~\ref{fig:wick}.

\begin{figure}[t]
	\centering
	\includegraphics[width=0.99\linewidth]{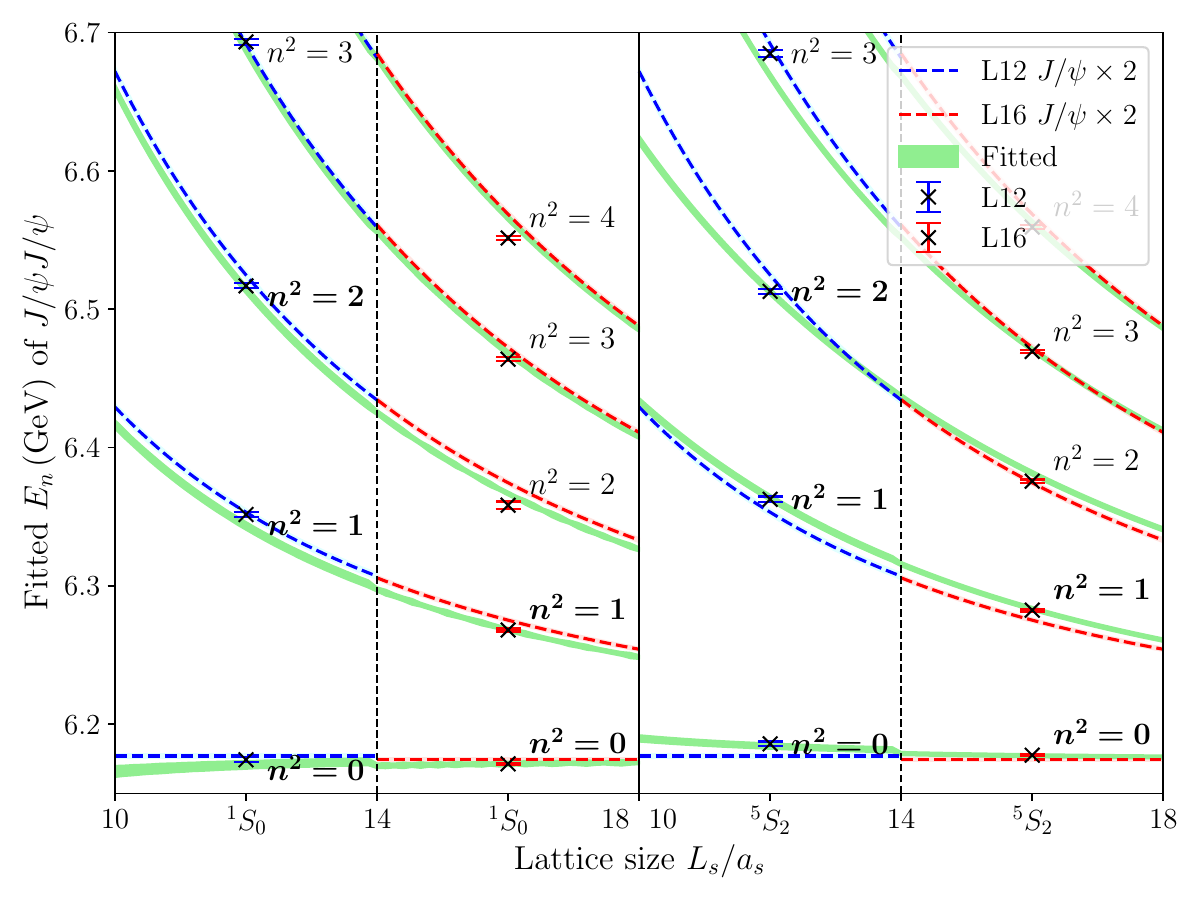}
	\caption{Finite volume energies in the ${^1}S_0$ (left block) and ${^5}S_2$ (right block) $J/\psi J/\psi$ channels at $m_\pi\approx 420$~MeV. 
		Blue and red dashed curves represent the noninteracting energies as functions of $L_s$. 
		Blue and red points are the lattice results on L12 and L16, respectively. 
		Green bands represent the theoretical predictions by the derived scattering amplitude $t(s)$.}
	\label{fig:fit-spectrum}
\end{figure}

After careful data analyses on the calculated correlation matrices and multiple self-consistency checks, we obtain the FVEs of the $^1S_0$ and $^5S_2$ $J/\psi J/\psi$ systems on the L12 and L16 lattices at each of the two pion masses (see details in the longer companion paper~\cite{Li:2025vbd}). 
Figure~\ref{fig:fit-spectrum} illustrates the FVEs (labeled by $E_n$) of the  $^1S_0$ (left) and $^5S_2$ $J/\psi J/\psi$ (right) systems at $m_\pi\approx 420~\rm{MeV}$ (the results at $m_\pi\approx 250$~MeV have a similar feature), where the points are the lattice data and the dashed lines show the noninteracting $J/\psi J/\psi$ energies $E_n^{(0)}$. 
It is seen that the deviation of $E_n$ from $E_n^{(0)}$ is significant in some energy regions, indicating sizable interactions.

\textit{Scattering analysis.---}
For the single channel $S$-wave $J/\psi J/\psi$ scattering, L\"{u}scher's quantization condition relates FVEs ($E_n$) to the phase shift $\delta_0$ via the relation 
\begin{equation}\label{eq:luscher}
	k\cot\delta_0(k)=\frac{1}{\pi L_s} \sum\limits_{\vec{n}\in Z^3} \frac{1}{\vec{n}^2-q^2},
\end{equation}
where $k=\sqrt{\frac{s}{4}-m_{J/\psi}^2}$ is the scattering momentum calculated at the invariant mass squared $s=E_n^2$ and $q^2 = (L_s/2\pi)^{2} k^2$ has been defined. 
The $S$-wave scattering amplitude $t(s)$ is subsequently obtained as
\begin{equation}\label{eq:pole}
	t(s)=\rho^{-1}(s) \frac{1}{\cot \delta_0(k)-i}=\frac{\sqrt{s}}{k\cot\delta_0(k)-ik},
\end{equation}
where $\rho=k/\sqrt{s}$ is the phase space factor of $J/\psi J/\psi$.

\begin{figure}[t]
	\centering
	\includegraphics[width=0.49\linewidth]{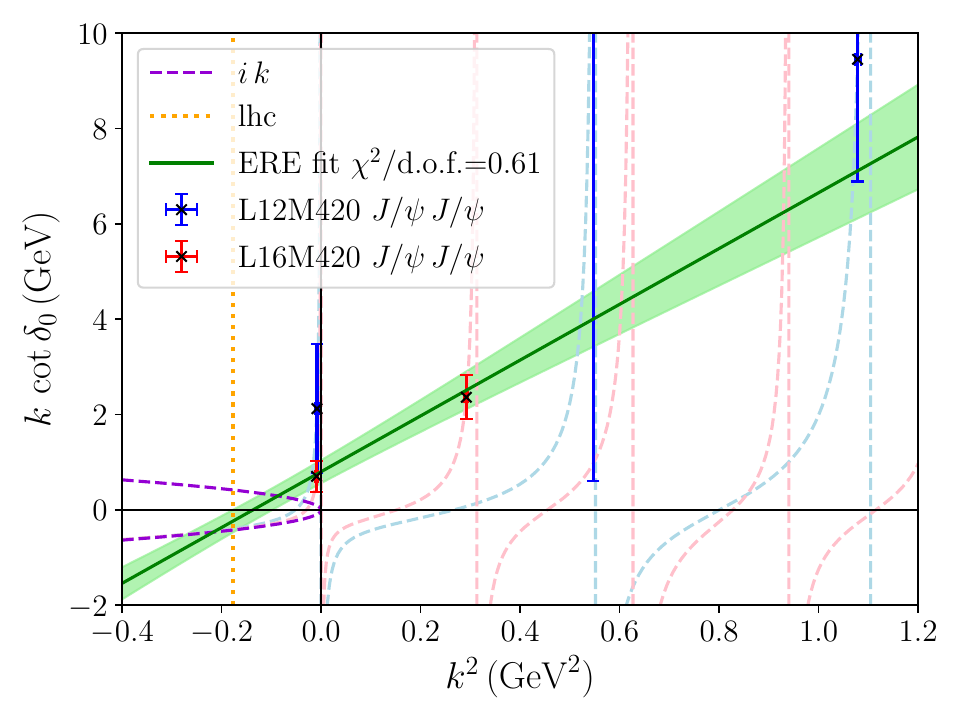}
	\includegraphics[width=0.49\linewidth]{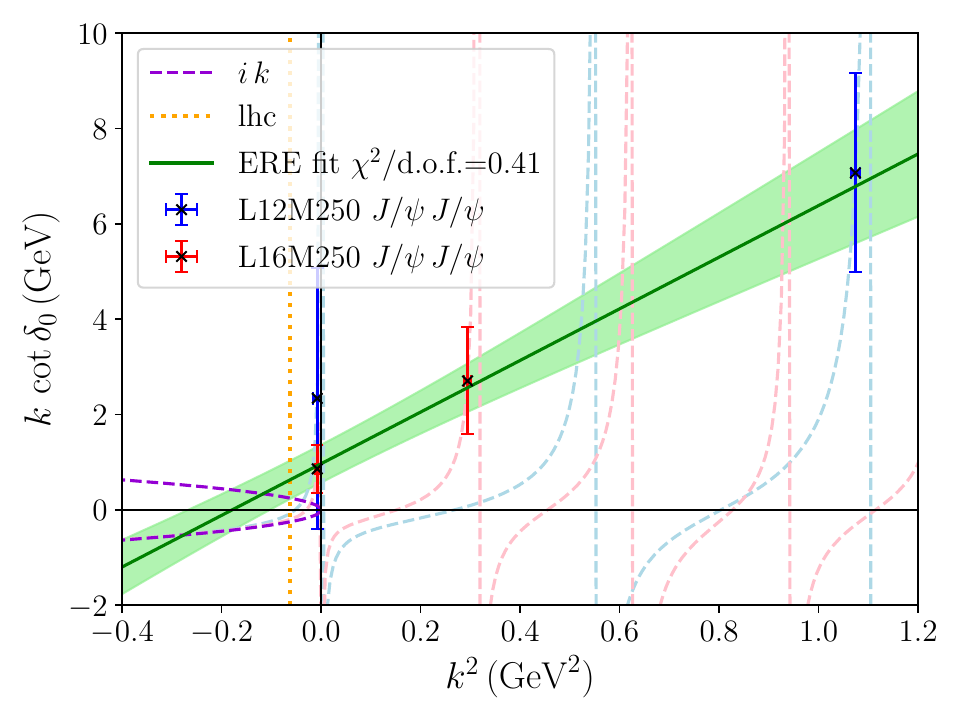}
	\caption{The ${}^1S_0$ $J/\psi J/\psi$ scattering phases $k\cot\delta_0(k)$ on M420 (left) and M250 (right).
		Blue and red points represent the results from FVEs on L12 and L16, respectively. 
		Dashed light blue~(red) curves illustrate Eq.~\eqref{eq:luscher}, the green band shows the ERE fit, and purple dashed line is the function $ik$ for $k^2<0$.
        The orange dotted line denotes the lhc induced by $\sigma$ exchange. }
	\label{fig:kcot_psi}
\end{figure}

The phase shift $k\cot\delta_0(k)$ of the ${}^1S_0$ channel is plotted as data points in Fig.~\ref{fig:kcot_psi} (the left panel for M420 and the right one for M250). 
The data exhibit a linear behavior up to $k^2 \simeq 1.2~\mathrm{GeV}^2$ and are described by the lowest effective range expansion (ERE) $k\cot\delta_0(k)=\frac{1}{a_0}+\frac{1}{2}r_0 k^2$ (green bands) with parameters
\begin{equation}\label{eq:erek2}
\begin{aligned}
&{\rm M420:~}a_0= 0.25(7) ~{\rm fm}, ~~ r_0=2.31(33) ~{\rm fm},\\
&{\rm M250:~}a_0= 0.20(8) ~{\rm fm}, ~~ r_0=2.14(41) ~{\rm fm},
\end{aligned}
\end{equation} 
where $a_0$ and $r_0$ are the scattering length and effective range, respectively.
The positive $a_0$ indicates a near-threshold attractive interaction in the ${}^{1}S_0$ channel. 
If analytically continued to the deep $k^2<0$ region, the ERE curve would intersect with the $ik$ curve (purple dashed line) twice and signal the pole singularities of $t(s)$ according to Eq.~\eqref{eq:pole}. 
However, the possible light meson ($\sigma$ or two pions) exchange will lead to a left-hand cut (lhc) starting from $(k^{\sigma}_{\mathrm{lhc}})^{2} = - m_{\sigma}^{2}/4\sim -1/m_\pi^2$ (indicated by the vertical orange dotted line in each panel of Fig.~\ref{fig:kcot_psi}), which sets an upper bound to the validity of ERE in the $k^2<0$ region~\cite{Du:2023hlu,Meng:2023bmz,Raposo:2023oru}. 
Since lhc alters the behavior of $k\cot\delta_0(k)$ drastically when $k^2$ approaches the lhc point~\cite{Du:2023hlu,Meng:2023bmz}, we do not take the interaction positions of the ERE and the $ik$ curves seriously, although for M420 the intersection point closer to the threshold resides to the right of the lhc point and may hint at a virtual state pole. 
A proper statement can be that the interplay of the lhc and the attractive interaction permits the existence of a scalar structure close to the $J/\psi J/\psi$ threshold. 

It is interesting to note that phenomenological studies reported the $X(6200)$ state (either $0^{++}$ or $2^{++}$) by analyzing the $J/\psi J/\psi$ mass spectra of LHCb, ATLAS and CMS. 
These studies used the $J/\psi J/\psi-J/\psi \psi(2S)$ coupled channel model or the $J/\psi J/\psi-J/\psi \psi(2S)-J/\psi \psi(3770)$) coupled channel model~\cite{Dong:2020nwy,Song:2024ykq}. 
Regardless of the theoretical assumptions, the model parameters, which are fitted from the experimental data, result in the near-threshold structure $X(6200)$, which can be a virtual state, a narrow resonance, or a bound state. 
If $X(6200)$ exists, our results favor its $0^{++}$ assignment because of the near-threshold repulsive interaction in the $^5S_2$ ($2^{++}$) $J/\psi J/\psi$ channel (see below).   

\begin{figure}[t]
	\centering
	\includegraphics[width=0.49\linewidth]{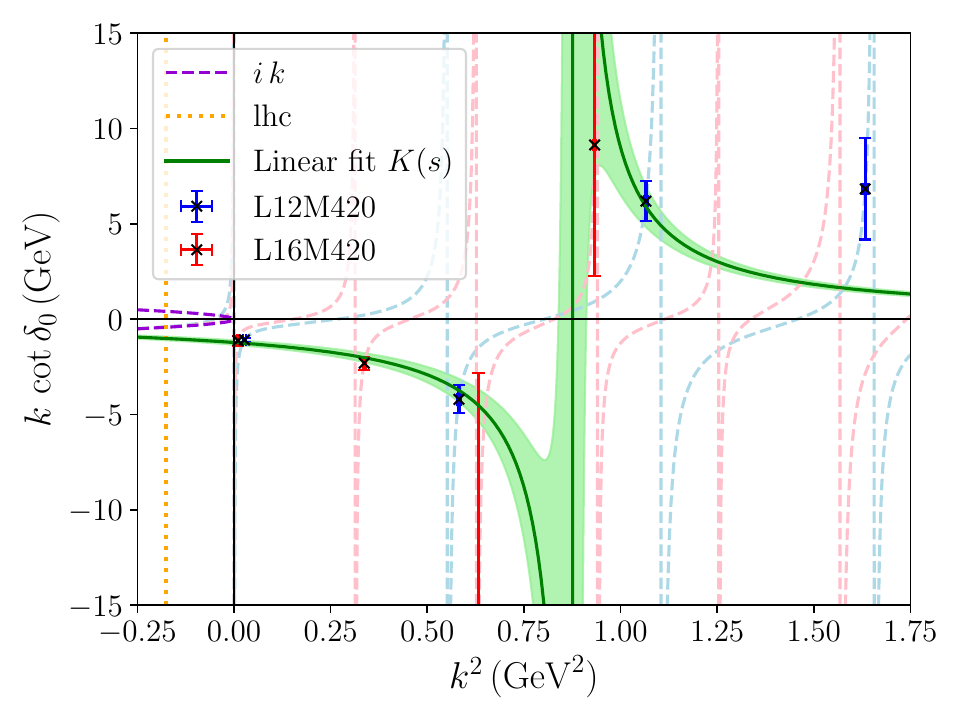}
	\includegraphics[width=0.49\linewidth]{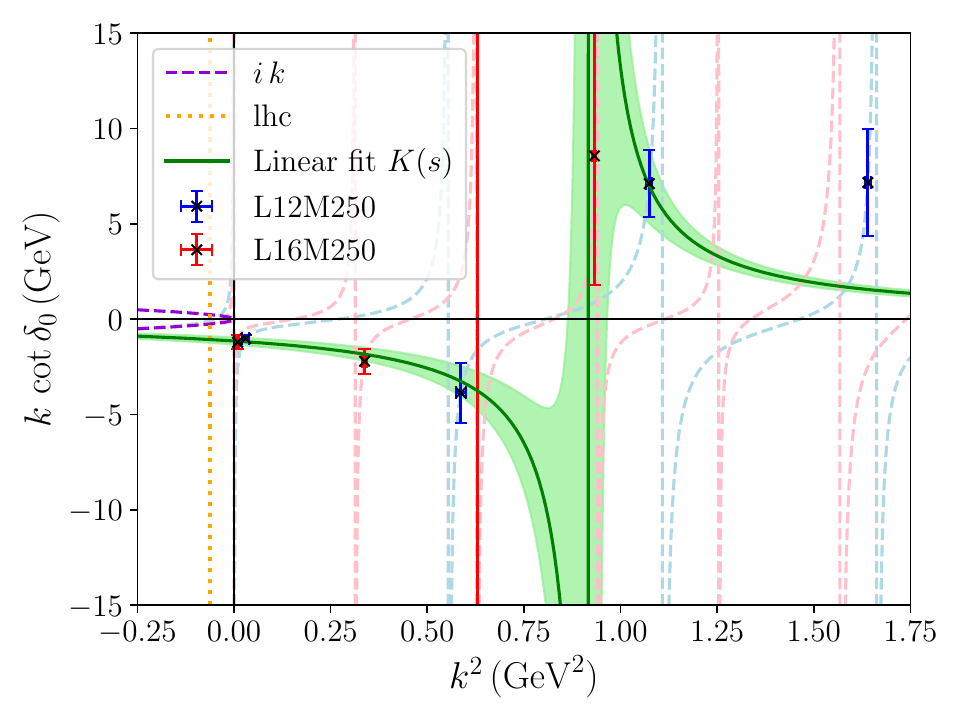}\\
	\includegraphics[width=0.49\linewidth]{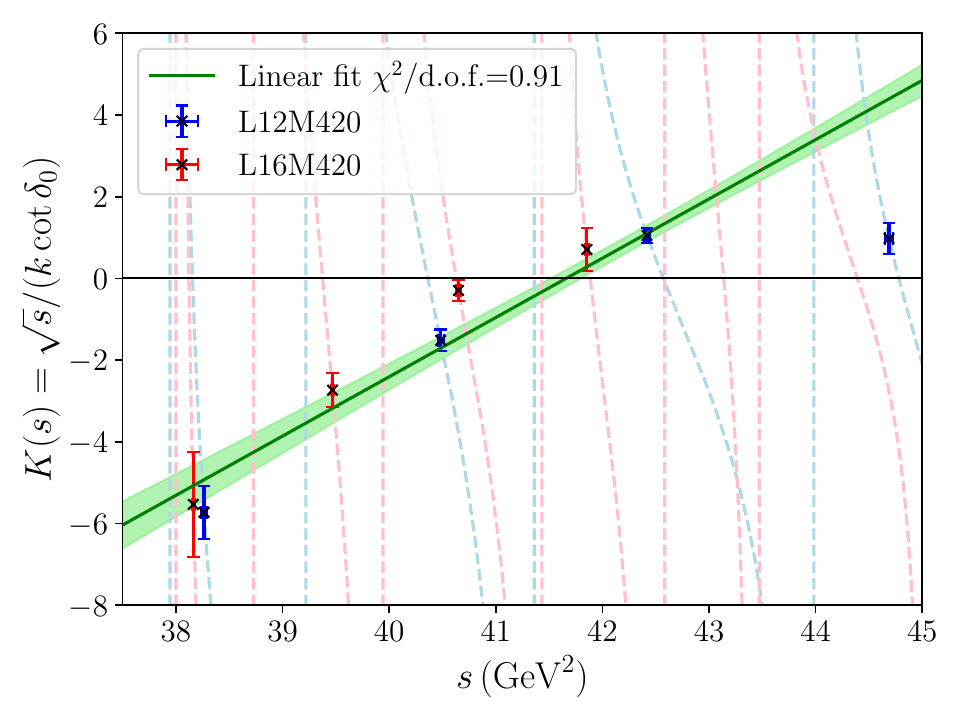}
	\includegraphics[width=0.49\linewidth]{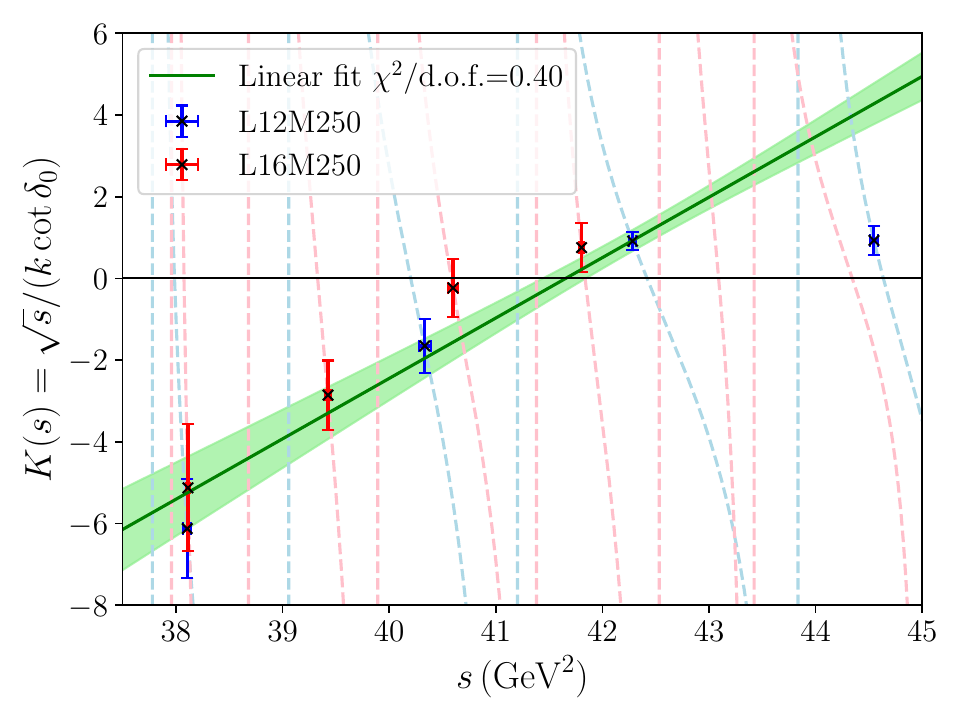}\\
	\includegraphics[width=0.49\linewidth]{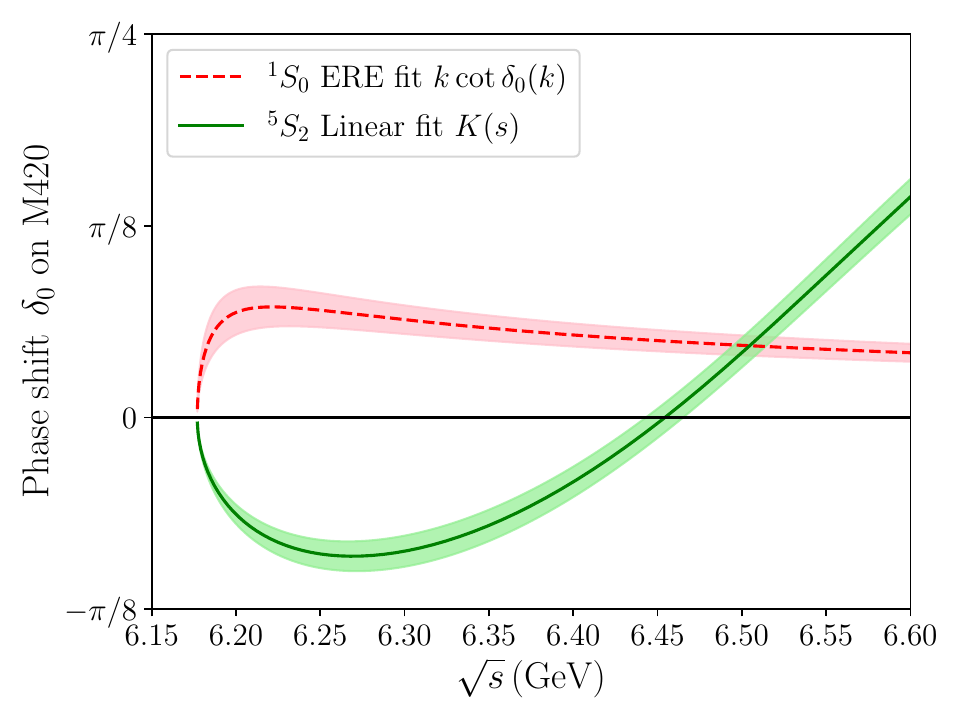} 
	\includegraphics[width=0.49\linewidth]{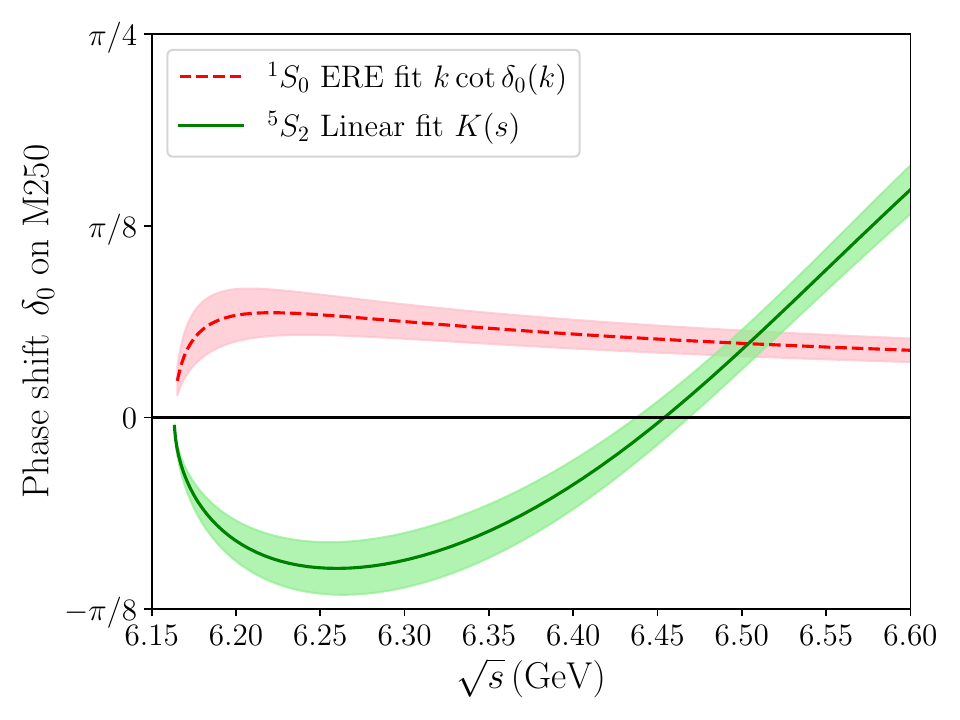}
	\caption{Scattering amplitudes of ${}^5S_2$ $J/\psi J/\psi$ scattering on M420 (left) and M250 (right). 
    {\bf Top}: $k \cot \delta_0(k)$ versus $k^2$. {\bf Middle}: $K(s)$ versus $s$. In each panel, blue and red points are lattice results from L12 and L16 ensembles, respectively, and the green band shows the fit up to $k^2\sim 1.25~{\rm GeV}^2$ or $s\sim 43~{\rm GeV}^2$ using Eq.~\eqref{eq:linear}.
    Note that the fits use only the three leftmost blue points and the two leftmost red points, which are considered to be reliable.
    {\bf Bottom}: phase shifts $\delta_0$ using the fitted parameters in both ${}^1S_0$ (red) and ${}^5S_2$ (green) $J/\psi J/\psi$ channels.}
	\label{fig:kcot_jpsi}
\end{figure}

The phase shift of the ${}^5S_2$ $J/\psi J/\psi$ scattering is shown in Fig.~\ref{fig:kcot_jpsi} (left column for M420 and right one for M250). 
The values of $k\cot\delta_0(k)$ (upper two panels) show a singular behavior around $k^2\sim 0.9~\rm{GeV}^2$ and cannot be described by a polynomial function. 
Thus, we adopt the $K$-matrix parametrization for $t(s)$, namely,
\begin{equation}\label{eq:pole-s}
	t(s)=\frac{K(s)}{1-i\rho(s)K(s)},\quad K(s)=\frac{\sqrt{s}}{k\cot\delta_0(k)},
\end{equation}
which obeys the single-channel unitarity. 
As shown in the middle row of Fig.~\ref{fig:kcot_jpsi}, $K(s)$ is approximately linear in $s$ up to $s\approx 43~{\rm GeV}^2$ (the two red points at $s\approx 41, 42~\rm{GeV}^2$ and the farthest blue point at $s\approx 45~\rm{GeV}^2$ are unreliable, as discussed in Ref.~\cite{Li:2025vbd}) and is described by 
\begin{equation}\label{eq:linear}
	K(s)=a ~ s + b.
\end{equation}
The fitted values of $a$ and $b$ are listed in Table~\ref{tab:parameters}. 
We note that these parameters almost reproduce the FVEs following Eq.~\eqref{eq:pole-s}, as illustrated by the green bands in the right block of Fig.~\ref{fig:fit-spectrum}. 

The scattering length $a_0 ({}^5S_2)$ can be determined in terms of $a$ and $b$ via the relation $\frac{1}{a_0}=\lim\limits_{k\to 0} k\cot\delta_0(k)=\frac{ 2m_{J/\psi}}{4a m_{J/\psi}^2+b}$. 
At the two $m_\pi$ values, we obtain 
\begin{equation*} 
	\begin{aligned}
		{\rm M420}:~&a_0({}^5S_2)=-0.163(16) ~{\rm fm},\\ 
		{\rm M250}:~&a_0({}^5S_2)=-0.171(29) ~{\rm fm},
	\end{aligned} 
\end{equation*}
which are consistent with $a_0=-0.165(16)$~fm in Ref.~\cite{Meng:2024czd}. 
The negative $a_0$ implies a near-threshold repulsive interaction in the ${}^5S_2$ $J/\psi J/\psi$ channel. 
Notably, $K(s)$ [thereby $t(s)$ and $\delta_0(s)$ according to Eq.~\eqref{eq:pole-s}] has a zero point at $s_0=-b/a$, namely, $\sqrt{s_0}=6.454(11)$~GeV at $m_\pi\approx 420$~MeV and 6.454(15)~GeV at $m_\pi\approx 250~\rm{MeV}$, as shown in the middle and bottom rows of Fig.~\ref{fig:kcot_jpsi}.
This zero on the positive real axis of $s$ is usually referred to as the Castillejo-Dalitz-Dyson~(CDD) zero~\cite{Castillejo:1955ed,Oller:2024lrk}. 
There are theoretical arguments suggesting that CDD zeros are closely related to discrete bare states in a free theory~\cite{Dyson:1957rgq,Krivoruchenko:2010ft,Li:2021cue}. 
Interestingly, phenomenological models predict that the lightest $2^{++}$ fully charmed tetraquark has a mass around 6.5~GeV~\cite{Chen:2016jxd,Lu:2020cns,Wang:2021kfv,Huang:2024jin}.

By solving the pole equation $1 - i~\rho(s)~K(s) = 0$ in the complex $k$ plane,
we obtain a pair of complex poles $\pm |{\rm Re}(k_R)| - i~|{\rm Im}(k_R)|$ corresponding to a resonance. (Another imaginary pole at $i|k_I|$ lies far from the threshold and is unphysical due to the lhc.)
With the values of $k_R$ in Table~\ref{tab:parameters}, the resonance parameters $\sqrt{s_R}=m_R-i\Gamma_R/2$ are determined to be
\begin{equation*}
	\begin{aligned}
		{\rm M420}:~\sqrt{s_R}&=(6.543(10) - \frac{i}{2} 0.548(34))~{\rm GeV},\\
		{\rm M250}:~\sqrt{s_R}&=(6.538(13) - \frac{i}{2} 0.537(56))~{\rm GeV}.
	\end{aligned}
\end{equation*}
Accordingly, the physical pole coupling $c_R$ is determined from the residue of the scattering amplitude $t(s) \approx \frac{|c_R|^2}{s_R - s}$ near the pole $s = s_R$, and the values of $|c_R|^2$ at the two $m_\pi$ are listed in Table~\ref{tab:parameters}. 
Actually, the behavior of $\delta_0$ (green bands in bottom panels of Fig.~\ref{fig:kcot_jpsi} already indicates the existence of this resonance. 
Here, $\delta_0(\sqrt{s})$ goes from zero to a negative value due to the near-threshold repulsive interaction and then goes up rapidly to pass zero (CDD zero). 
The large slope of $\delta_0(\sqrt{s})$ around the CDD zero $\sqrt{s_0}$ signals an additional attractive interaction~\cite{Krivoruchenko:2010ft} and is also a typical feature of a resonance. 
In contrast, the $^1S_0$ phase shift (red bands) is smooth in the large $\sqrt{s}$ region apart from the rapid rise near the threshold because of the attractive interaction.
\begin{figure}[t]
	\centering
	\includegraphics[width=0.9\linewidth]{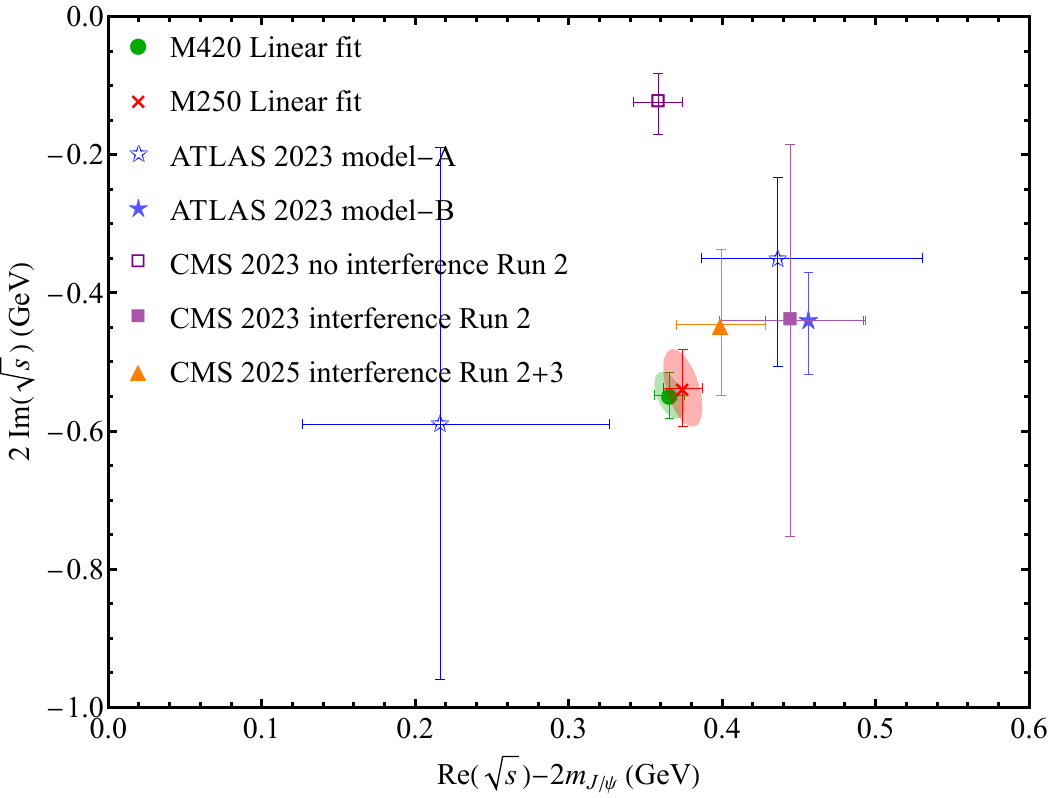}
	\caption{Pole positions in the complex $\sqrt{s}$ plane with a linear fit at two pion masses (green dot for M420 and red cross for M250).
    The experimental measurements from ATLAS~\cite{ATLAS:2023bft} and CMS~\cite{CMS:2023owd, CMS:2025fpt, CMS:2025xwt} are represented by blue and purple (orange) points, respectively.}
	\label{fig:pole-exp}
\end{figure}

\begin{table}[t]
	\caption{Fitting and resonance~($R$) parameters.}
	\label{tab:parameters}
	\begin{ruledtabular}
		\begin{tabular}{lrr}
			Parameters & $m_\pi\approx 420~{\rm MeV}$ & $m_\pi\approx 250~{\rm MeV}$\\
			\hline
			$a$~(GeV$^{-2}$)                       &1.45(12)        &1.48(20)        \\
			$b$                                    &-60.4(4.9)      &-61.6(8.4)       \\
			\hline
			$\operatorname{Re}(k_R)$~(GeV)         &1.140(17)       &1.148(24)        \\
			$\operatorname{Im}(k_R)$~(GeV)         &-0.393(20)      &-0.383(34)       \\
			$m_R$~(GeV)                            &6.543(10)       &6.538(13)        \\
			$\Gamma_R$~(GeV)                       &0.548(34)       &0.537(56)        \\
			$|c_R|^2$~(GeV$^2$)                    &16.2(0.6)        &15.9(1.1)       \\
		\end{tabular}
	\end{ruledtabular}
\end{table}

We compare our resonance pole positions (green point for M420 and red point for M250) with those of $X(6600)$ and $X(6400)$ reported by ATLAS~\cite{ATLAS:2023bft} (blue points) and CMS~\cite{CMS:2023owd,CMS:2025fpt,CMS:2025xwt} (purple and orange points) in Fig.~\ref{fig:pole-exp}, where the $\mathrm{Re}(\sqrt{s})$ axis is shifted by $2m_{J/\psi}$ to eliminate the slight difference between our lattice value of $m_{J/\psi}$ and the physical $m_{J/\psi}$.
In the 6.2–6.7~GeV region, ATLAS identified either two interfering $S$-wave resonances, namely, $(m, \Gamma) \approx (6.41(11), 0.59(39))$~GeV for $X(6400)$ and $(6.63(8), 0.35(14))$~GeV for $X(6600)$ using their model A, or a single resonance $X(6600)$ with $(m, \Gamma) \approx (6.65(3), 0.44(7))$~GeV via model B.
Using Run 2 data, CMS observed the $X(6600)$ state with parameters $(m,\Gamma)\approx(6.552(15),0.124(30))~\mathrm{GeV}$ in a fit with three noninterfering resonances, or $(6.638(48),0.440(300))~\mathrm{GeV}$ when interference effects were included. 
Notably, CMS recently performed a combined analysis of Run 2 and Run 3 data, classifying $X(6600)$, $X(6900)$, and $X(7100)$ as members of a $2^{++}$ $T_{cc\bar c\bar c}$ tetraquark family~\cite{CMS:2025fpt, CMS:2025xwt}.
Our lattice QCD results are found to be compatible with the properties of the $X(6600)$.

\begin{figure}[t]
	\centering
	\includegraphics[width=0.49\linewidth]{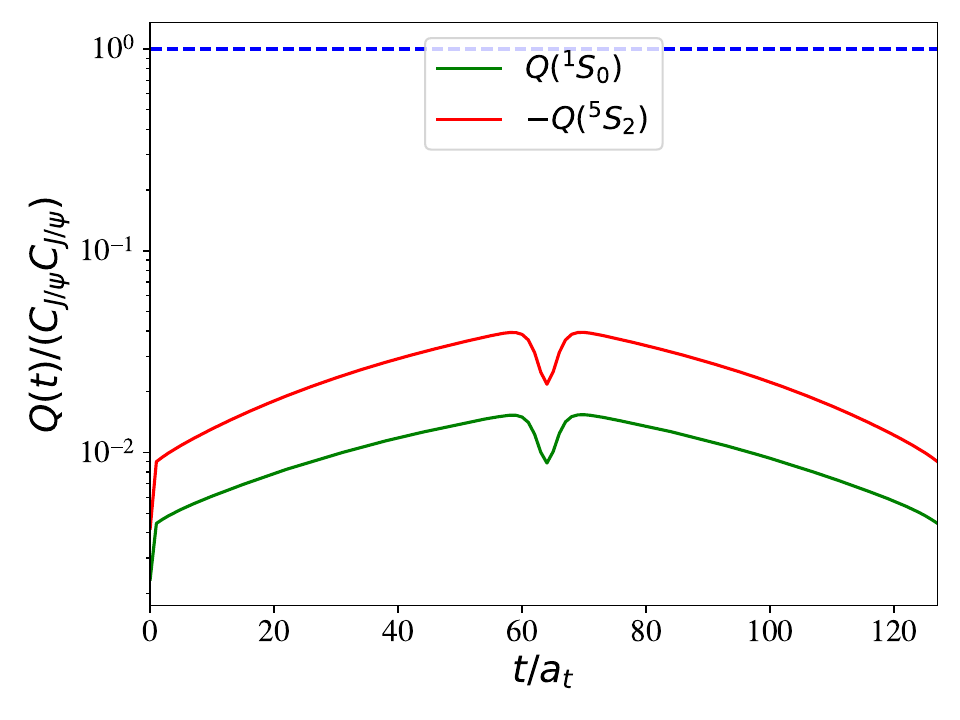}
	\includegraphics[width=0.49\linewidth]{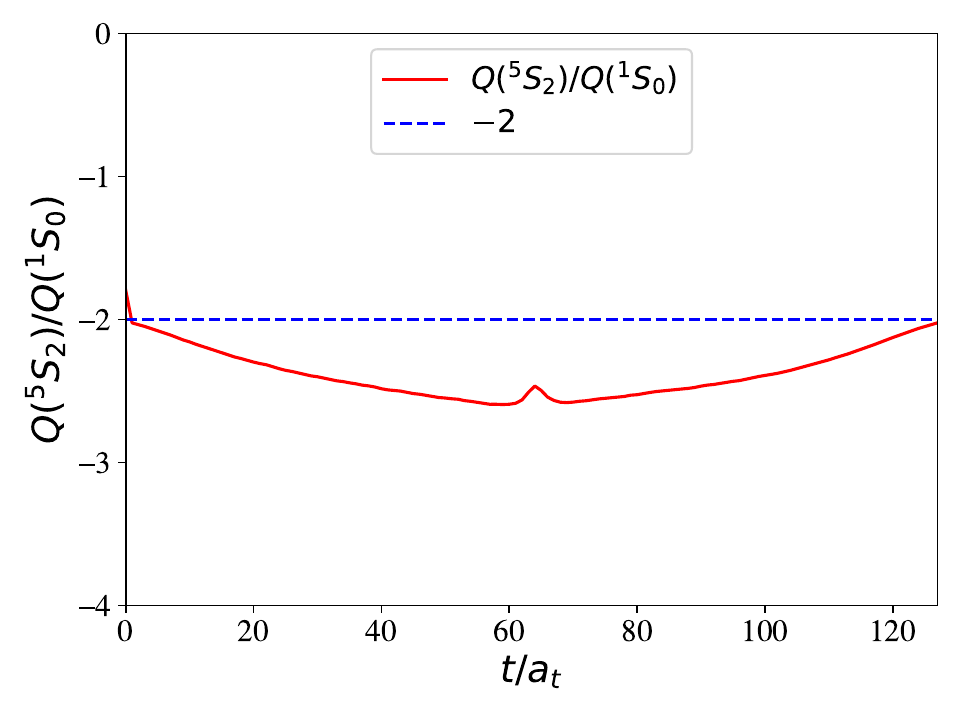}
	\caption{Ratio of the quark rearrangement~(Q) components in the (${}^5S_2$,~${}^1S_0$) $J/\psi J/\psi$ systems. All curves are for the lowest $C(n^2=0,~t)$ case. }
	\label{fig:diagrams}
\end{figure}

\textit{Near threshold $J/\psi J/\psi$ interaction.---}
In order to understand the observation that the near-threshold $J/\psi J/\psi$ interaction is attractive in the ${}^1S_0$ channel and repulsive in the ${}^5S_2$ channel, we investigate the underlying dynamics in a qualitative way. 
Let $C(0,t)$ be the correlation function of the $J/\psi J/\psi$ operator $\mathcal{O}_{\psi\psi}(n^2=0)$, which mainly contains the information of the lowest $S$-wave $J/\psi J/\psi$ state in the large $t$ region. 
The lattice calculation of $C(0,t)$ considers diagrams~(a) and (b) in Fig.~\ref{fig:wick}. 
If the contributions from diagrams~(a) and (b) are denoted by $D(t)$ and $Q(t)$, respectively, we have $C(0,t)=D(t)+Q(t)$. 
The $Q(t)$ term reflects the quark-rearrangement (QR) effect during the propagation of $J/\psi J/\psi$, while $D(t)$ can be viewed as the propagation of two free $J/\psi$'s plus the $t$-channel light-flavor (gluons or light mesons) exchange effects. 
Since the light-flavor exchange effect is much weaker than that of QR~\cite{Li:2025vbd}, $D(t)$ is temporarily approximated by the squared $J/\psi$ propagator $C_\psi^2(t)$. 
As such the relative contribution of $Q(t)$ to $C(0,t)$ is $Q(t)/C_\psi^2(t)\approx \epsilon e^{-\delta_Q t}$, as shown in the left panel of Fig.~\ref{fig:diagrams}, where one can easily read out a tiny negative $\delta_Q$ and $|\epsilon|\sim \mathcal{O}(10^{-2})$ with $\epsilon<0$ for $^1S_0$ (green) and $\epsilon>0$ for $^5S_2$ (red). 
Thus, we have the estimate of the minimal $J/\psi J/\psi$ energy, $E_\mathrm{min}\approx 2m_{J/\psi}+\epsilon \delta_Q$~\cite{Chen:2022vpo,Li:2025vbd}, which gives $E_\mathrm{min}({}^1S_0) < 2m_{J/\psi}$ (attractive interaction) and $E_\mathrm{min}({}^5S_2) > 2m_{J/\psi}$ (repulsive interaction). 

The opposite signs of $Q(t)$ in these two channels are understood as follows.
Quark rearrangement proceeds through the exchange of at least one gluon, as depicted in diagram~(b) of Fig.~\ref{fig:wick}. 
In the near-threshold scattering, the two interchanging charm (anti)quarks are expected to preserve their spin orientations. 
Let $|m_s\rangle_i~(m_s=0,\pm)$ denote the spin states of the $i$th $J/\psi$ in the $J/\psi J/\psi$ system; then, after QR, the spin configuration of the initial $J/\psi J/\psi$ state changes as 
\begin{equation}
	\begin{aligned}
		&|0\rangle_1 |0\rangle_2  \overset{\mathrm{QR}}{\longrightarrow} \frac{1}{2}\left(|0\rangle_1 |0\rangle_2+|+\rangle_1 |-\rangle_2+|-\rangle_1|+\rangle_2\right),\\ 
		&|\pm\rangle_1 |\mp\rangle_2 \overset{\mathrm{QR}}{\longrightarrow} \frac{1}{2}|0\rangle_1 |0\rangle_2, ~~~ |\pm\rangle_1 |\pm\rangle_2  \overset{\mathrm{QR}}{\longrightarrow} |\pm\rangle_1 |\pm\rangle_2.
	\end{aligned}
\end{equation}
If $H_{\rm QR}$ is the interaction Hamiltonian for QR, we have the following relation:
\begin{equation}
	r_\mathrm{QR}=\frac{\langle {}^5S_2|H_\mathrm{QR}|{}^5S_2\rangle}{\langle {}^1S_0|H_\mathrm{QR}|{}^1S_0\rangle}\approx -2,
\end{equation}
where $|^1S_0\rangle=\frac{1}{\sqrt{3}} \left( |+\rangle_1 |-\rangle_2 + |-\rangle_1 |+\rangle_2 - |0\rangle_1 |0\rangle_2 \right)$ and $|^5S_2\rangle=|+\rangle_1 |+\rangle_2$ are the spin wave functions  (based on the rotational invariance). 
This relation is supported quantitatively by the numerical result that $\frac{Q(t, {}^5S_2)}{Q(t, {}^1S_0)} \sim -2$, as shown in the right panel of Fig.~\ref{fig:diagrams} (the deviation from $-2$ comes mainly from the difference of $\delta_Q$ in the two channels).

\textit{Summary.---}
We investigate the $^1S_0$ and $^5S_2$ $J/\psi J/\psi$ scattering with center-of-mass energies up to 6.6~GeV using $N_f=2$ lattice QCD ensembles at $m_\pi\approx 250~\rm{MeV}$ and $420~\rm{MeV}$. 
In the ${}^0S_1$ $J/\psi J/\psi$ channel, the observed near-threshold attractive interaction permits the existence of a $0^{++}$ near-threshold structure like $X(6200)$.
In the $^5S_2$ channel, the $J/\psi J/\psi$ system exhibits a repulsive interaction near the threshold, while the scattering amplitude has a resonance pole associated with a CDD zero around 6.45~GeV. 
The parameters of this tensor resonance are determined to be $(m_R,\Gamma_R)=(6.543(10), 0.548(34))~\rm{GeV}$ for $m_\pi\approx 420~\rm{MeV}$ and $(6.538(13),0.537(56))~\rm{GeV}$ for $m_\pi\approx 250~\rm{MeV}$, which are compatible with those of $X(6600)$ [or $X(6400)$] observed by ATLAS and CMS, and they support the $2^{++}$ assignment of $X(6600)$ by the recent CMS study. 
Furthermore, the difference of the $J/\psi J/\psi$ interactions in the $^1S_0$ and $^5S_2$ channels can be explained neatly by the dominance of the quark rearrangement effects. 
Although our results show little sensitivity to the pion masses used in this work, the systematic uncertainties associated with unphysical quark masses, as well as finite-volume and finite-lattice-spacing effects, should be addressed in future lattice studies with more sophisticated setups.
On the other hand, the origin of the CDD zero and the formation of the tensor resonance remain to be understood theoretically.

\textit{Acknowledgments.}
We thank Qiang Zhao and Feng-Kun Guo for insightful discussions. 
This work is supported by the National Key Research and Development Program of China (Grant No.~2025YFB3003602) and the National Natural Science Foundation of China (Grants No.~12293060, No.~12293065, No.~12293061, No.~12205311, and No.~11935017). 
W.S. and G.L. are supported by the Chinese Academy of Sciences (Grant No.~YSBR-101), and G.L. is also supported by the China Postdoctoral Science Foundation (Grant No.~2025M773362). 
We acknowledge the use of Chroma software~\cite{Edwards:2004sx}, QUDA library~\cite{Clark:2009wm,Babich:2011np}, and PyQUDA package~\cite{jiang2024usequdalatticeqcd}. 
Computations were performed on HPC clusters at the Institute of High Energy Physics (Beijing), China Spallation Neutron Source (Dongguan), and the ORISE computing environment.

\bibliography{reference}

\end{document}